\documentclass[preprint,aps,showpacs]{revtex4}
\usepackage{amsmath}
\usepackage{graphicx}
\usepackage{dcolumn}
\usepackage{bm}

\begin{document}

\title{Continuous phase transition of a fully frustrated 
XY model in three dimensions}

\author{Kwangmoo Kim and David Stroud}
\affiliation{Department of Physics, The Ohio State University, Columbus,
Ohio 43210, USA}
\date{\today}
\begin{abstract}
We have used Monte Carlo simulations, combined with finite-size
scaling and two different real-space renormalization group
approaches, to study a fully frustrated three-dimensional XY model
on a simple cubic lattice. This model corresponds to a lattice of
Josephson-coupled superconducting grains in an applied magnetic
field ${\bf H}=(\Phi_{0}/a^2)(1/2,1/2,1/2)$.   We find that the
model has a continuous phase transition with critical temperature
$T_c = 0.681J/k_{\mathrm{B}}$, where $J$ is the XY coupling
constant, and critical exponents $\alpha/\nu = 0.87\pm 0.01$,
$v/\nu = 0.82\pm 0.01$, and $\nu = 0.72\pm 0.07$, where
$\alpha$, $v$, and $\nu$ describe the critical behavior of the
specific heat, helicity modulus, and correlation length.  We briefly
compare our results with other studies of this model, and with a
mean-field approximation.
\end{abstract}


\maketitle

\section{\label{sec:level1}Introduction}

The classical XY model has been widely used for decades as a model
for phase transitions in materials with interacting spins.  The
dimensionality $n$ of the spins ($n = 2$ for the XY model) is
independent of the lattice dimensionality $d$, which
can be either $2$ or $3$ for physically relevant systems.  The
$d = 2$ XY model undergoes the well-known Kosterlitz-Thouless (KT)
transition \cite{kt}, characterized by an unbinding of
vortex-antivortex pairs at the KT transition temperature $T_{\mathrm{KT}}$.
The KT transition is a continuous phase transition, but with unique
critical properties \cite{kt}.  The $d = 3$ XY model exhibits a more
conventional phase transition with well-known critical
exponents \cite{li1,gottlob2}. It is thought to describe many
ferromagnetic materials with two-component spins. In addition,
it describes phase transitions in which the ``spins'' actually
represent the phases of a complex order parameter, such as the
$\lambda$ transition in He$^{4}$ and superconductor-to-normal
phase transition at zero applied magnetic field.

The {\em frustrated} classical XY model, in either $d = 2$ or
$d = 3$, has a much wider range of phase diagrams than does the
unfrustrated case just mentioned.  In this model, the coupling
between spins is such that the ground state of the system cannot
minimize all the bond energies simultaneously.  Interest in this
model was greatly increased when it was realized that this model
described real systems, such as Josephson junction arrays in an
applied magnetic field.   The first demonstration that the fully
frustrated XY model undergoes a continuous phase transition in $d =
2$ was given by Teitel and Jayaprakash \cite{teitel1}, using Monte
Carlo (MC) techniques.  This work was later extended to other values
of the so-called frustration parameter $f$ \cite{teitel2}, leading
to an extensive literature on two-dimensional (2D) frustrated XY 
model on various lattices and at different values of 
$f$ \cite{cataudella,gottlob,dukovski}.

The $d = 3$ frustrated XY model has also been studied extensively,
in part because it is believed to describe flux line lattice (FLL)
melting \cite{nelson1} under an applied magnetic field in high-T$_c$
superconductors. Hetzel {\it et al.} \cite{hetzel} used a uniformly
frustrated 3D XY model on a stacked triangular lattice to study the
melting of an unpinned Abrikosov lattice in a type-II
superconductor.   They showed convincingly that this melting
transition is first order, rather than continuous --- a prediction
subsequently confirmed by experiment.
Earlier work on a frustrated XY model on a simple cubic lattice,
with magnetic field parallel to one of the lattice axes \cite{shih1}
found a continuous phase transition.  Li and Teitel \cite{li2,li3}
used a uniformly frustrated XY model similar to that of Ref.\ \cite{hetzel}
to calculate the properties of the vortex line liquid which appears
above the melting temperature, and to investigate the possibility of
a further phase transition between an entangled and a disentangled
vortex line liquid above the FLL melting transition. Chen and
Teitel \cite{chen} later extended this work to the more realistic
case of uniaxially anisotropic couplings; they suggested that
there was another phase transition temperature $T_{cz}$ above the
FLL melting temperature $T_m$, where superconducting coherence
parallel to the applied magnetic field would vanish.
Chin {\it et al.} \cite{chin} have, however, suggested
that this apparent existence of a transition to a disentangled vortex
liquid is a result of finite system sizes and simulation times.
More recently, the uniformly frustrated 3D XY model was studied by
Monte Carlo methods on a simple cubic lattice with a magnetic field
parallel to the $[111]$ direction \cite{hwang}.  For this choice of
field direction, the simple cubic lattice behaves as a stack of 2D
triangular lattices with $ABCABC\cdots$ stacking.   It was found
that this system, similar to that studied in Ref.\ \cite{hetzel}, 
exhibits a clear first-order FLL melting transition.   Nguyen and 
Sudb\o\ \cite{nguyen1} considered a uniformly frustrated anisotropic 
Villain model (an approximation to the uniformly frustrated XY model) 
to study the phase diagram of a uniaxially anisotropic high-T$_c$
superconductor as a function of the applied magnetic field and
temperature.   They found two phase transitions: the
lower-temperature one is FLL melting, while that at higher
temperature involves the destruction of the phase coherence in the
direction of the applied magnetic field.

In this paper, we study phase transitions in the fully frustrated XY
model in $d = 3$, using primarily MC simulations. This
model is of interest, in part, because it may be relevant to FLL
melting in the high-T$_c$ materials.  In addition, because of great
advances in microfabrication techniques, it is now possible to make
microscale or nanoscale arrays of Josephson superconducting grains 
{\it in three dimensions}.   Such an array should, in a suitable 
applied magnetic field, be describable by a frustrated XY model, 
at least to a first approximation.  It has also recently been 
suggested that a fully frustrated XY model might also be realized 
by an assembly of cold atoms on a suitably constructed optical 
lattice \cite{polini}.

The fully frustrated XY model is characterized by the frustration
vector ${\bf f} = (1/2, 1/2, 1/2)$, as further defined below.  Such a
model has been previously studied by Diep {\it et al.} \cite{diep},
who found that, in contrast to the ${\bf f} = (1/3, 1/3, 1/3)$ case
studied in Ref.\ \cite{hwang}, there was a continuous phase
transition.   We extend the work of Ref.\ \cite{diep} by calculating
the critical behavior of the helicity modulus, equivalent to the
spin-wave stiffness constant (or to the superfluid density in a
superconductor).  We also carry out a more extensive finite-size
scaling analysis than done by those workers, thus obtaining more
accurate information about the critical behavior. Our results do,
however, confirm that the phase transition is continuous, not
first order.

The remainder of this paper is organized as follows. In Sec.\ 
\ref{sec:level2}, we present the formalism for calculating the
thermodynamic properties and the critical exponents of our model.
In Sec.\ \ref{sec:level3}, we give the results of our Monte
Carlo calculations, finite-size scaling methods, and renormalization
group methods. Section \ref{sec:level4} presents a summary and discussions.

\section{\label{sec:level2}Formalism}

We now describe the model Hamiltonian on which our calculations are
carried out.  For convenience, we present this Hamiltonian as it
applies to a simple cubic lattice of superconducting grains in the
presence of an applied external magnetic field ${\bf H}$, though the
model is not limited to this application, of course.  Then the
frustrated XY model in $d = 3$ is described by the Hamiltonian
\begin{equation}
\mathcal{H}=-\sum_{\langle ij\rangle}J_{ij}\cos
(\phi_{i}-\phi_{j}-A_{ij}). \label{eq:hamil}
\end{equation}
Here  $\phi_{i}$ is the phase of the superconducting order parameter
on the $i$th site, $A_{ij}=(2\pi/\Phi_{0})\int_{i}^{j}{\bf A}\cdot
d{\bf l}$, $\Phi_{0}=hc/2e$ is the flux quantum, ${\bf A}$ is the
vector potential, $J_{ij}$ is the coupling constant between the
$i$th site and $j$th site, and $\langle ij \rangle$ represents a sum
over all distinct pairs of nearest-neighbor sites on a simple cubic
lattice. We assume a constant coupling between each nearest-neighbor
pair of sites, so $J_{ij}=J$ and $J  > 0$; we neglect the possible
dependence of $J$ on the applied magnetic field ${\bf H}$ and the
temperature $T$.  We also assume weak screening as in Ref.\
\cite{hwang}.  Thus, the local magnetic field ${\bf B}$ is
approximated by the applied magnetic field ${\bf H}$.

The $x$ component of the frustration vector ${\bf f}=(f_x,f_y,f_z)$
is defined by
\begin{equation}
\sum_{p}{\textstyle^{(x)}}A_{ij}=2\pi f_{x},
\label{eq:plaq}
\end{equation}
where the sum is taken along the sides of a plaquette on the $yz$
plane of the lattice; analogous definitions hold for $f_y$ and
$f_z$.   If the simple cubic lattice has lattice constant $a$,
$f_{i}=B_{i}a^2/\Phi_{0}$ represents the flux through a single
plaquette perpendicular to the $i$th axis, in units of one flux
quantum.

One can also define the vortex number (or {\it vorticity}) of each
square plaquette.  For example, the vorticity of a plaquette lying
in the $yz$ plane is defined by
\begin{equation}
n_{x}=f_{x}+\frac{1}{2\pi}\sum_{p}{\textstyle^{(x)}}(\phi_{i}-\phi_{j}-A_{ij}),
\end{equation}
where the sum runs counterclockwise around the perimeter of the
plaquette, viewed from the positive $x$ direction, and the phase
differences are chosen so that $0 \leq \phi_{i} < 2\pi$.
$n_{y}$ and $n_{z}$ are defined analogously.

In order to study possible phase transitions within this model, we
have calculated several quantities.  One of these is the specific
heat $C_V$ per site, given by
\begin{equation}
C_{V}=\frac{\langle \mathcal{H}^2 \rangle - \langle \mathcal{H}
\rangle^2}{Nk_{\mathrm{B}}T^2},
\end{equation}
where $N$ is the total number of sites in the lattice, $\mathcal{H}$
is the Hamiltonian in Eq.\ (\ref{eq:hamil}), and $\langle \cdots \rangle$
denotes an average within the canonical ensemble.   A first-order
phase transition is generally indicated by a $\delta$-function-like
anomaly in $C_V$, while a continuous phase transition is signaled by
lattice-size-dependent divergence in the $C_V$.

To study the vortex lattice melting, we calculate a suitable
vorticity density-density correlation function.  Specifically, we
first introduce the Fourier transform of the vorticity density 
$n_i({\bf k}) = \sum_{\bf R}n_i({\bf R})\exp(i{\bf k}\cdot{\bf R})$,
where $n_i({\bf R})$ represents the vorticity of a plaquette
centered at ${\bf R}$ and oriented perpendicular to the $i$th
axis.  We then calculate some of the correlation functions
\begin{equation}
g_{ij}({\bf r})=\frac{1}{N^2 f_{i}f_{j}}\sum_{{\bf R}, {\bf
R}^\prime}{\textstyle ^{\prime}} \langle n_{i}({\bf R})n_{j}({\bf R}^\prime)
\rangle,
\end{equation}
where the sum is carried out over all ${\bf R}$ and ${\bf R}^\prime$
such that ${\bf R}^\prime - {\bf R} = {\bf r}$.   In our actual
simulations, we use periodic boundary conditions in all three
directions.   In practice, it is more convenient to compute the
Fourier transform of $g_{ij}({\bf r})$.  This Fourier transform is
known as the vortex structure factor, and is given by $S_{ij}({\bf
k})=\sum_{\bf r} g_{ij}({\bf r})e^{i{\bf k}\cdot{\bf r}}$.  Because
of the periodic boundary conditions, $S_{ij}({\bf k})$ is defined
only for ${\bf k} = [2\pi/(N_xa)](m_1, m_2, m_3)$, where $m_i$ are
integers, and the computational cell is assumed to contain $N =
N_x^3$ sites.  We obtain $g_{ij}({\bf r})$ by first computing
$S_{ij}({\bf k})$ and then Fourier-transforming back into real
space.

The phase transition in this model is best characterized by the {\it
helicity modulus tensor} $\gamma_{\alpha \beta}$ \cite{fisher}.
$\gamma_{\alpha\beta}$ measures the stiffness of the phase $\phi$
against an external twist. It is defined as the second derivative of
the free energy with respect to an infinitesimal phase twist \cite{ohta}
\begin{equation}
\gamma_{\alpha \beta}=\frac{1}{N}\left. \frac{\partial^{2}F}{\partial
\delta_{\alpha}\partial\delta_{\beta}}\right |_{\delta=0}.
\end{equation}
$\gamma_{\alpha\beta}$ is conveniently calculated by adding a
fictitious vector potential ${\bf A}^\prime$ to the Hamiltonian in Eq.\
(\ref{eq:hamil}), as
\begin{equation}
\gamma_{\alpha \beta}=\frac{1}{N}\left.
\frac{\partial^{2}F}{\partial A_{i}^\prime \partial A_{j}^\prime}
\right |_{{\bf A}^\prime=0},
\end{equation}
in the presence of periodic boundary conditions on the phases.  The
resulting diagonal components are readily evaluated, with the result
\begin{eqnarray}
\gamma_{\alpha \alpha} & = & \frac{1}{N} \left\langle
\sum_{\langle ij \rangle} J_{ij}\cos (\phi_{i}-\phi_{j}-A_{ij})(\mbox{\^{e}}_{ij}
\cdot\mbox{\^{e}}_{\alpha})^{2} \right\rangle \nonumber \\
& - &\frac{1}{Nk_{B}T}\left\langle \left[\sum_{\langle ij \rangle}J_{ij}
\sin (\phi_{i}-\phi_{j}-A_{ij})(\mbox{\^{e}}_{ij}\cdot
\mbox{\^{e}}_{\alpha})\right]^{2}\right\rangle \nonumber \\
& + &\frac{1}{Nk_{B}T}\left\langle \sum_{\langle ij \rangle}J_{ij}
\sin (\phi_{i}-\phi_{j}-A_{ij})(\mbox{\^{e}}_{ij}\cdot
\mbox{\^{e}}_{\alpha})\right\rangle^{2}.
\end{eqnarray}
Here $\mbox{\^{e}}_{ij}$ is the unit vector from the $i$th site to
the $j$th site, and $\mbox{\^{e}}_{\alpha}$ is the unit vector in
the $\alpha$ direction.   Since the helicity modulus $\gamma_{\alpha
\alpha}$ is proportional to the $\alpha$ component of the superfluid
density, it follows that when $\gamma_{\alpha \alpha}>0$, there is
nonzero phase coherence in the $\alpha$ direction, and when
$\gamma_{\alpha\alpha} \rightarrow 0$, the phase coherence is lost.
Therefore, the superconductor-to-normal phase transition occurs at
the temperature at which $\gamma_{\alpha\alpha} \rightarrow 0$.

The correlation time $\tau$ is a measure of how long it takes for
the system to lose its memory of its previous state.  $\tau$ can be
obtained by calculating a time-displaced autocorrelation function
\cite{newman4}. The time-displaced autocorrelation function
$\chi_{E}(t)$ of the energy, for example, is defined by
\begin{eqnarray}
\chi_{E}(t) & = & \int [E(t^\prime)-\langle E \rangle][E(t^\prime+t)-\langle E
\rangle]\mathrm{d}t^\prime \nonumber \\
& = & \int[E(t^\prime)E(t^\prime+t)-\langle E \rangle^2]\mathrm{d}t^\prime,
\label{eq:auto}
\end{eqnarray}
where $t$ and $t^\prime$ are two different Monte Carlo ``times.''
Since Monte Carlo simulations involve fictitious dynamics, these
times are not related in any obvious way to a physical time.
However, the length of the Monte Carlo time does give a measure of
the persistence of various MC states in time.

Equation\ (\ref{eq:auto}) can be expressed more conveniently when we
have a set of measurements of the energy $E(t)$ from $t = 0$
(after equilibration) up to some maximum time $t_{\max}$.
In this case, Eq.\ (\ref{eq:auto}) becomes
\begin{eqnarray}
\chi_{E}(t) & = & \frac{1}{t_{\max}-t}\sum_{t^\prime=0}^{t_{\max}-t}
E(t^\prime)E(t^\prime+t) \nonumber \\ & - & \frac{1}{(t_{\max}-t)^2}
\sum_{t^\prime=0}^{t_{\max}-t}E(t^\prime)\sum_{t^\prime=0}^{t_{\max}-t}
E(t^\prime+t).
\end{eqnarray}
Since the autocorrelation function is expected to fall off
exponentially at long times as
\begin{equation}
\chi_{E}(t) \sim e^{-t/\tau},
\end{equation}
the correlation time $\tau$ can be calculated from
\begin{equation}
\int_{0}^{\infty}\frac{\chi_{E}(t)}{\chi_{E}(0)}\mathrm{d}t
= \int_{0}^{\infty}e^{-t/\tau}\mathrm{d}t=\tau.
\end{equation}
This expression for $\tau$ is also called {\it integrated
correlation time}.   MC ``measurements'' will be statistically
independent only if they are separated by intervals of $\sim 2\tau$
or more. In addition, the correlation time $\tau$ is related to the
dynamic exponent $z$ by
\begin{equation}
\tau \sim \xi^{z}\sim L^{z},
\end{equation}
where $\xi$ is the {\em correlation length} (equal to $L$ at the
$T_c(L)$). The dynamic exponent measures the extent of the critical
slowing down. The smaller $z$ is, the more accurate are the numerical
measurements.

If there is a continuous phase transition, various quantities should
exhibit singular behavior near the transition temperature $T_c$.  If
we define a reduced temperature by
\begin{equation}
t=\frac{T-T_{c}}{T_{c}},
\end{equation}
then in the thermodynamic limit the correlation length $\xi$, the
specific heat per site $C_{V}$, and the helicity modulus $\gamma$
near $T_c$ are expected to vary as
\begin{eqnarray}
\xi & \sim & |t|^{-\nu}, \label{eq:xi} \\
C_{V} & \sim & |t|^{-\alpha}, \label{eq:cv} \\
\gamma & \sim & |t|^{v} \label{eq:gamma},
\end{eqnarray}
where $\nu$, $\alpha$, and $v$ are critical exponents.   If $T_c$ is
known, the simulation data can be fitted to the expected asymptotic
form to obtain values of the critical exponents.  But since $T_c$ is
typically not known in advance, it is usually more accurate to use a
different method.   One such method is finite-size scaling, which
extracts values for the critical exponents by investigating how
measurements depend on the size $L$ of the system \cite{newman2}.
This procedure is carried out by expressing a quantity of interest
in terms of the correlation length and then introducing a new
dimensionless function, known as a {\it scaling function}.  For
example, $C_V$ and $\gamma$ are expressed as
\begin{eqnarray}
C_{V}(L,t) & = & L^{\alpha/\nu}\tilde{C_{V}}(L^{1/\nu}t), \\
\gamma(L,t) & = & L^{-v/\nu}\tilde{\gamma}(L^{1/\nu}t),
\label{eq:gammasf}
\end{eqnarray}
where $L  = N_xa$ is the linear system dimension.   Since the
scaling functions $\tilde{{C}_V}$ and $\tilde{\gamma}$ should depend
on a single variable, we can make all the data for each system size
$L$ fall on the same curve by appropriately adjusting the values of
the critical exponents and $T_c$. When this happens, we assume that
we have the correct values for these quantities.

Another method is the phenomenological renormalization group (PRG)
\cite{barber,nightingale,li1}.  To calculate the critical behavior
of $\gamma$, for example, we consider two different system sizes $L$
and $L^\prime$ and introduce the ratio
\begin{equation}
P_{\gamma}(L,L^\prime,t)\equiv \frac{\gamma(L,t)}{\gamma(L^\prime,t)}.
\label{eq:prg}
\end{equation}
From Eq.\ (\ref{eq:gammasf}) this ratio becomes
$P_{\gamma}=(L/L^\prime)^{-v/\nu}$ when $t=0$. Therefore, if one plots two
different curves of $P_{\gamma}(L,L^\prime,t)$ versus $t$ with the same
ratio of $L/L^\prime$, the temperature at which they intersect is $T_c$,
and the value of $P_\gamma$ at that temperature yields the ratio
$v/\nu$.

The critical temperature $T_{c}$, or its inverse value $K_{c}=J/
(k_{\mathrm{B}}T_{c})$, and the critical exponent $\nu$ can also be
determined from Binder's fourth-order cumulant $U_{L}$ \cite{binder}
defined by
\begin{equation}
U_{L}=1-\frac{\langle s^{4}\rangle}{3\langle s^{2}\rangle^{2}},
\end{equation}
where $s=(1/N)\sqrt{(\sum_{i}^{N}\cos \phi_{i})^{2}
+(\sum_{i}^{N}\sin \phi_{i})^{2}}$.   The scaling form for the
Binder's cumulant is $U_L = \tilde{U}(L^{1/\nu}t)$, without any
prefactor. Hence, $U_L$ can be Taylor expanded about $T_{c}$ as
\begin{eqnarray}
U_{L} & = & U_{0}+U_{1}L^{1/\nu}\left(1-\frac{T}{T_{c}}\right)+\cdots
\nonumber \\
& = & U_{0}+U_{1}L^{1/\nu}\left(1-\frac{K_{c}}{K}\right)+\cdots.
\label{eq:taylor}
\end{eqnarray}
If we plot $U_{L}$ for several values of $L$ as a function of
temperature or its inverse value, they will intersect at the
critical temperature $T_c$.   To obtain the exponent $\nu$, we can
calculate $(\mathrm{d}U_{L}/\mathrm{d}K)_{K = K_c}$, where
$K = J/(k_{\mathrm{B}}T)$. From Eq.\ (\ref{eq:taylor}), we find that
\begin{equation}
\left. \frac{\mathrm{d}U_{L}}{\mathrm{d}K}\right|_{K_{c}}
= \frac{U_1}{K_c}L^{1/\nu}.
\end{equation}
Hence, the ratio of the two slopes for different values of $L$ gives
$\nu$.

\section{\label{sec:level3}Monte Carlo Calculations}

To carry out our Monte Carlo calculations, we used the standard
Metropolis algorithm with periodic boundary conditions in all three
directions.  We started with a random phase configuration at
temperature $T=1.0J/k_{\mathrm{B}}$, then cooled down to $T=0.4J/
k_{\mathrm{B}}$ in steps of $0.01J/k_{\mathrm{B}}$, except near
$T_c$, where we decreased the temperature in steps of
$0.005J/k_{\mathrm{B}}$. At each $T$, we took 50000 Monte Carlo
steps per site through the entire lattice for equilibration, and
then calculated the expectation values of the quantities of interest
by averaging over an additional 20000 MC steps.  We also used the
final configuration of the previous $T$ as the starting one of the
current $T$. Near $T_c$, the system undergoes critical slowing
down; so we increased the number of MC steps up to $5\times 10^5$
for equilibration and $2 \times 10^5$ for averaging.

Instead of considering continuous angles between $0$ and $2\pi$ for
the phases $\phi_i$ of the order parameter on each site, we used the
360-state clock model, which allows angles of $0^{\circ}$,
$1^{\circ}$, $2^{\circ}$,..., $359^{\circ}$.   This
simplification should have no effect on our results, since it is
known that there is no distinction between the continuum and the
discrete results for the $n$-state clock model when $n > 20$
\cite{thijssen}.

To calculate the phase factors $A_{ij}$, we use the gauge ${\bf A} =
(\Phi_0/a^2)(f_yz\hat{x}+f_zx\hat{y}+f_xy\hat{z})$, where the
frustration is ${\bf f} = f_x\hat{x} + f_y\hat{y} + f_z\hat{z}$.
Thus, for example, the phase factors $A_{ij}^{(z)}$ arising from the
field component parallel to $z$ all vanish except for bonds in the
$y$ direction; for these bonds, and $x = na$, $A_{ij}^{(z)}$ is
given by
\begin{equation}
A_{ij}^{(z)}=2\pi n f_{z}.
\end{equation}
The phase factors $A_{ij}^{(x)}$ and $A_{ij}^{(y)}$ are given by
analogous expressions.  For ${\bf f} = (1/2, 1/2, 1/2)$ and periodic
boundary conditions in all three directions, all the phase factors
are equal to either $0$ or $\pi$, and thus all the couplings are of
the form $\pm J\cos(\phi_i - \phi_j)$.  This choice of gauge
automatically satisfies the condition given in Eq.\ (\ref{eq:plaq}).

\begin{figure}[h]
\begin{center}
\includegraphics[width=\textwidth]{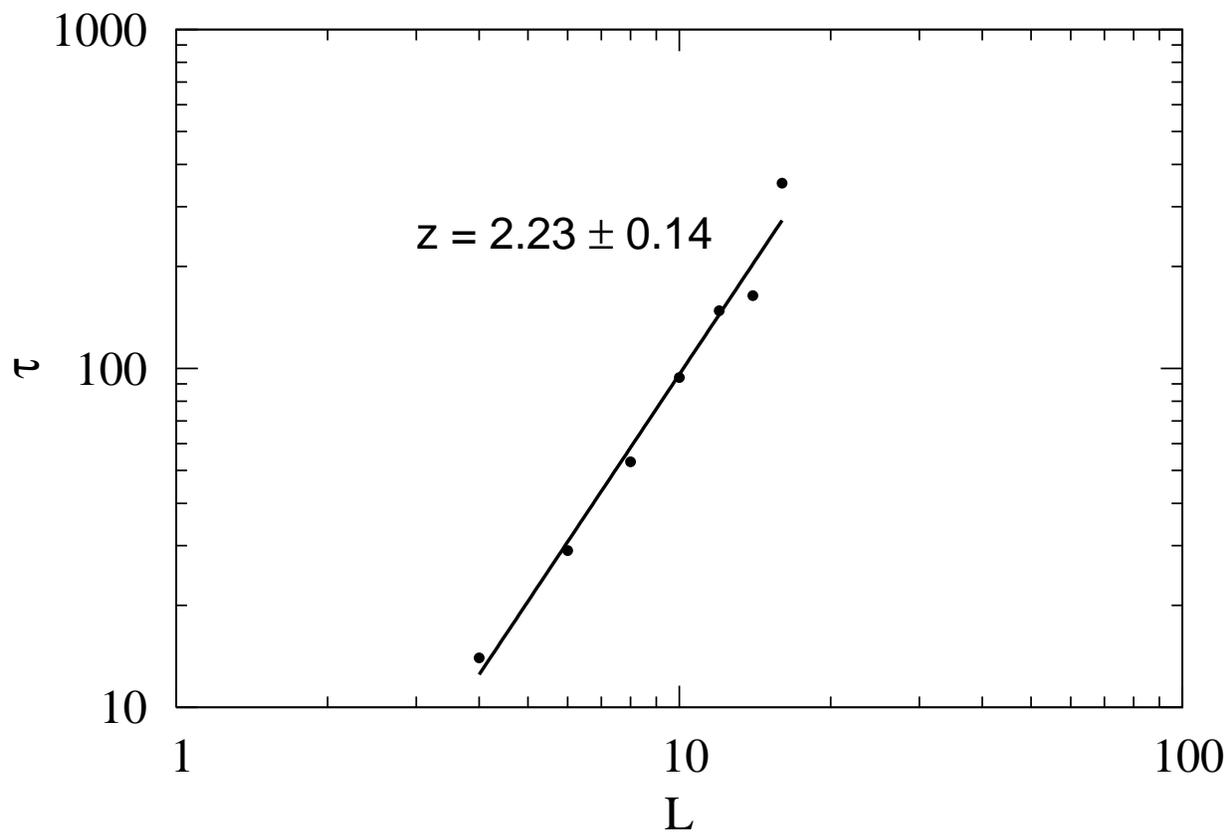}\\%
\end{center}
\caption{\label{fig:scaling_time}Log-log plot of the data in Table
I, corresponding to a fit of $\tau$ to the function $\tau = AL^z$.
The slope of the fitting line to the data yields the dynamic
exponent $z=2.23\pm 0.14$.}
\end{figure}

We turn now to our numerical results. The values of the correlation
time $\tau$ for several lattice sizes are given in Table I,
at the estimated transition temperature $T_c(L)$ for the fully
frustrated XY model [${\bf f} = (1/2, 1/2, 1/2)$] of size $L = N_xa$.
$\tau$ clearly increases monotonically with increasing lattice size.
From the fits to data in Fig.\ \ref{fig:scaling_time},
we get the dynamic exponent $z=2.23\pm 0.14$ in our system.

\begin{figure}[h]
\begin{center}
\includegraphics[width=\textwidth]{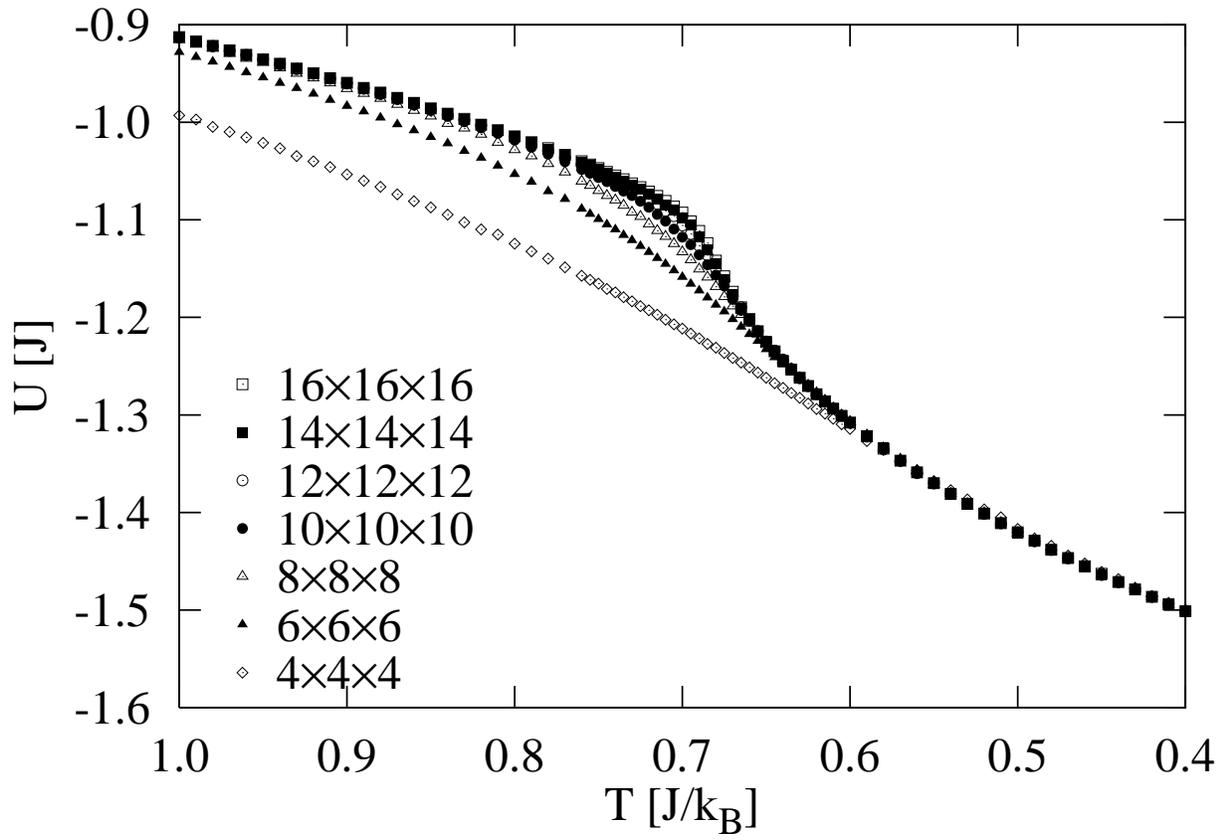}\\%
\end{center}
\caption{\label{fig:energy}Internal energy per site as a function of
temperature $T$ for several lattice sizes, as indicated.  Note that
$T$ decreases with increasing distance along the horizontal axis.}
\end{figure}

The internal energy per site $U\equiv E/N$, expressed in units of
the coupling constant $J$, is shown in Fig.\ \ref{fig:energy}. It is
quite size-dependent for $L < 10$ but quickly converges for
$L > 10$.   The sharp drop in $U$ near $T=0.7J/k_{\mathrm{B}}$
suggests but obviously does not prove the occurrence of a phase
transition near that temperature.

\begin{figure}[h]
\begin{center}
\includegraphics[width=\textwidth]{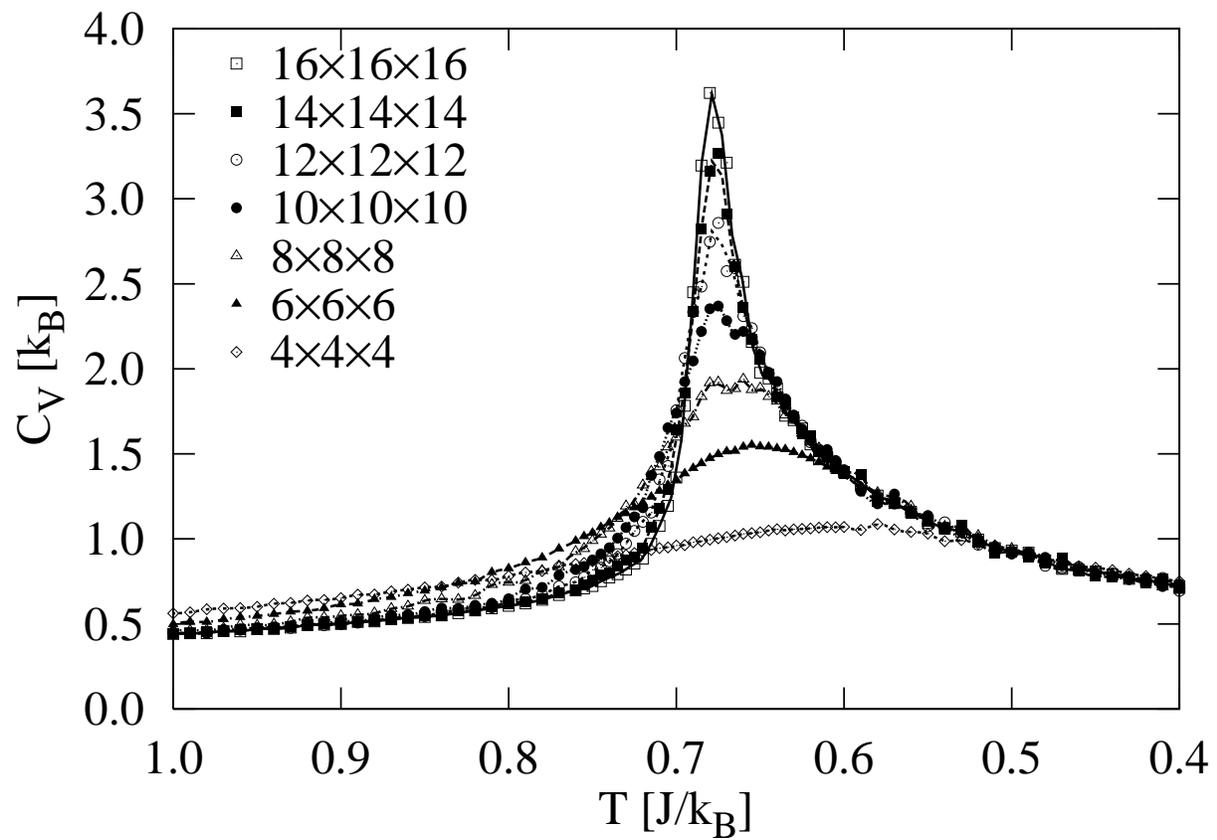}\\%
\end{center}
\caption{\label{fig:sh}The specific heat per site $C_V$ as a
function of temperature for several lattice sizes.  The lines are
cubic spline fits to the data.}
\end{figure}

The calculated specific heat per site $C_V$ is shown in Fig.\
\ref{fig:sh}.  There is a clear peak near $T \sim
0.7J/k_{\mathrm{B}}$ which becomes sharper with increasing $L$,
suggesting a continuous phase transition.  This behavior is similar
to that of the 2D fully frustrated XY model as in Ref.\
\cite{teitel1} and that of the ordered simple cubic lattice with
${\bf f}=(0,0,1/2)$, as in Ref.\ \cite{shih1}.   As in those two
examples, the finite magnitude of $C_V$ at its peak is a result of
the finite system size; otherwise, $C_V$ would diverge at $T_c$ in
the infinite system.

\begin{figure}[h]
\begin{center}
\includegraphics[width=\textwidth]{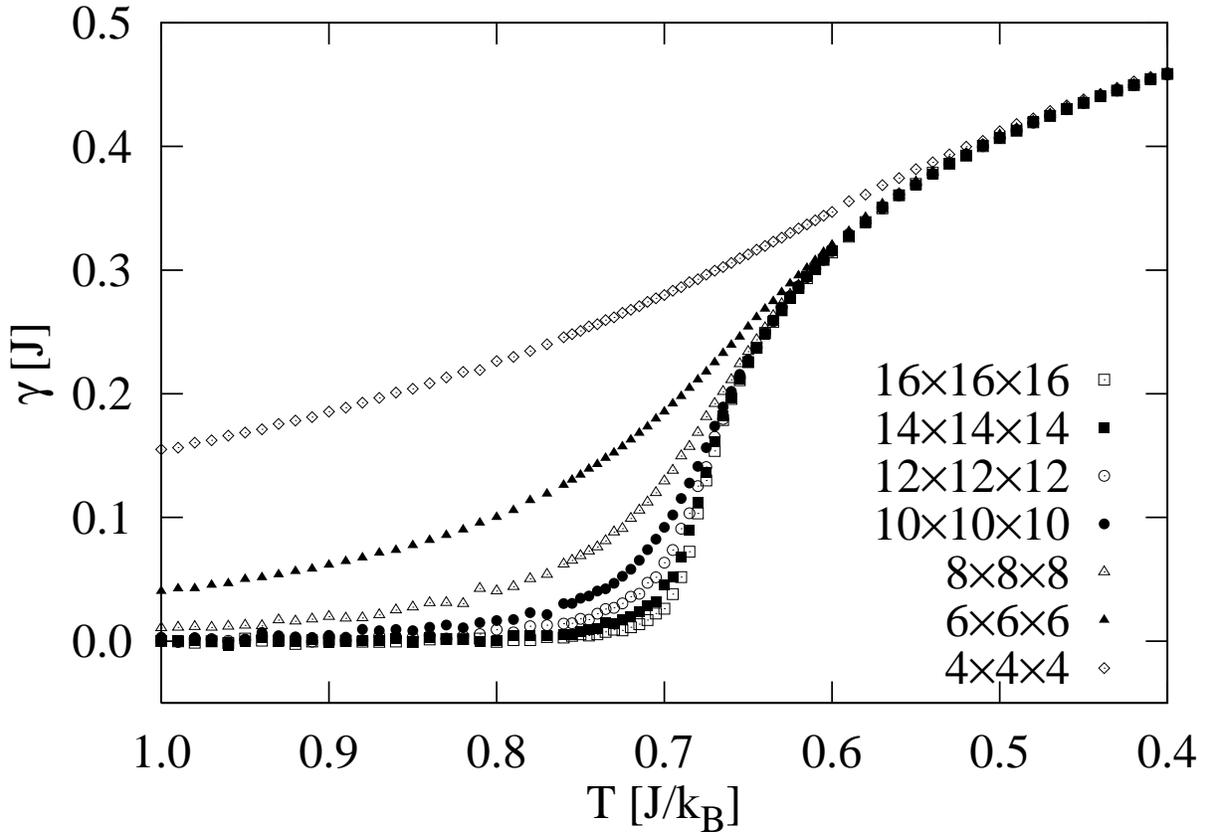}\\%
\end{center}
\caption{\label{fig:gamma}The averaged helicity modulus $\gamma =
(\gamma_{xx}+\gamma_{yy}+\gamma_{zz})/3$ as a function of
temperature for several lattice sizes.}
\end{figure}

Next, we turn to the behavior of the helicity modulus tensor.  Since
$\gamma_{xx} = \gamma_{yy} = \gamma_{zz}$ for this isotropic system,
we calculated the average $\gamma=(\gamma_{xx}+\gamma_{yy}+\gamma_{zz})/3$
in order to improve the statistics.  $\gamma(T)$ increases with decreasing
$T$, and shows a fairly clear drop to near zero near $T = 0.7 J/k_{\mathrm{B}}$
when $L > 10$, as can be seen in Fig.\ \ref{fig:gamma}. For $L = 10$
or smaller, $\gamma(T)$ shows a broad transition region.
A more accurate value for $T_{c}$ will be given further below.

To estimate the statistical errors in our calculation of
$\gamma(T)$, we used the jackknife method \cite{newman3}, which is
carried out as follows.   For each $T$, we made $n$ independent
numerical measurements of $\gamma(T)$, with measurements separated
by at least two correlation times. From these $n$ measurements, we
calculated a value $\gamma(T)$ for the helicity modulus.   Next, we
removed the first measurement from this set of $n$ measurements to
calculate the helicity modulus $\gamma_{1}$ with the remaining $n-1$
measurements. To calculate $\gamma_{2}$ we restored the first
measurement to the set and removed the second measurement, and so
on.  Thus, $\gamma_i$ is calculated with the $i$th measurement
removed from the set. The error estimate for $\gamma$ is given by
\begin{equation}
\sigma_{\gamma}=\sqrt{\sum_{i=1}^{n}(\gamma_{i}-\gamma)^2}.
\end{equation}
The error bars from this method are shown in Fig.\ \ref{fig:gamma},
but they are smaller than the symbol sizes.

In Fig.\ \ref{fig:densitytc}, we show the evolution of the internal
energy per site $U$ as a function of MC time at the
transition temperature $T_{c}=0.681J/k_{\mathrm{B}}$ for a lattice
size $12\times 12\times 12$. The three intensity plots for each $a$
and $b$ show the density-density correlation functions
$g_{zz}(x,y,L_{z}/2)$, $g_{xx}(L_{x}/2,y,z)$, and $g_{yy}(x,L_{y}/2,z)$
(denoted by $xy$, $yz$, and $zx$, respectively) of the vortices at
two different times as indicated in the energy evolution curve.
Each intensity plot is an average of one correlation time $\tau=148$
MC steps per site through the entire lattice.  The two times correspond
to energies slightly below and slightly above the average value and
they are separated by a very large MC time. In contrast to the results of
Ref.\ \cite{hwang}, there is no indication of a vortex lattice phase
in these plots; instead, both seem to show vortex liquid phases
(although there are partial latticelike formations, especially on
the $zx$ plane in window $a$).  This behavior suggests (as does the
diverging specific heat peak in Fig.\ \ref{fig:sh}) that the phase
transition at ${\bf f}=(1/2,1/2,1/2)$ is continuous, rather than
first order as at ${\bf f}=(1/3,1/3,1/3)$. However, at very low
temperatures ($T = 0.01 J/k_{\mathrm{B}}$), the correlation
functions show a clearer indication of an ordered phase, as shown
in Fig.\ \ref{fig:densitylowt}.  Specifically, we see evidence of
a checkerboard lattice in all three directions, but especially in
the $yz$ windows.

This ordering is clearer if we look at the vortex structure factor
$S_{zz}({\bf k})$ of the lattice.  Figure\ \ref{fig:densitysf} shows
this structure factor for ${\bf k}$ parallel to the magnetic field,
at several temperatures, including the two shown in Figs.\
\ref{fig:densitytc} and \ref{fig:densitylowt}.  Here
$k_{\|} = |{\bf k}|$ and ${\bf k} = [2\pi/(N_xa)](m, m, m)$, where
$m= 0$, $1$,..., $N_x - 1$.  At low temperatures, the vortex structure
factors at $k_{\|}=\sqrt{3}\pi/a$ are nonzero, but they do not increase 
monotonically with decreasing temperature.  Instead, they increase down 
to $T=0.30J/k_{\mathrm{B}}$, then decrease down to $T=0.01J/k_{\mathrm{B}}$.
We believe that this behavior may represent some kind of
``polycrystalline'' domain structure of the vortex at low temperatures.
As shown below, the system has several eigenmodes, and the exact
admixture of these eigenmodes may change as the temperature varies.
The insets to Fig.\ \ref{fig:densitysf} are the vortex density-density
correlation functions $g_{zz}(x,y,L_{z}/2)$ at two different
temperatures $T=0.30J/k_{\mathrm{B}}$ and $T=0.681J/k_{\mathrm{B}}$.
We can see a clear checkerboard pattern at $T=0.30J/k_{\mathrm{B}}$.
The vortex structure factor appears to go to zero near the same
temperature as $T_c$ where the helicity modulus vanishes, although we
did not collect enough numerical data to determine the
temperature dependence of the structure factor peak near $T_c$. This
point is discussed further below.

\begin{figure}[h]
\begin{center}
\includegraphics[width=\textwidth]{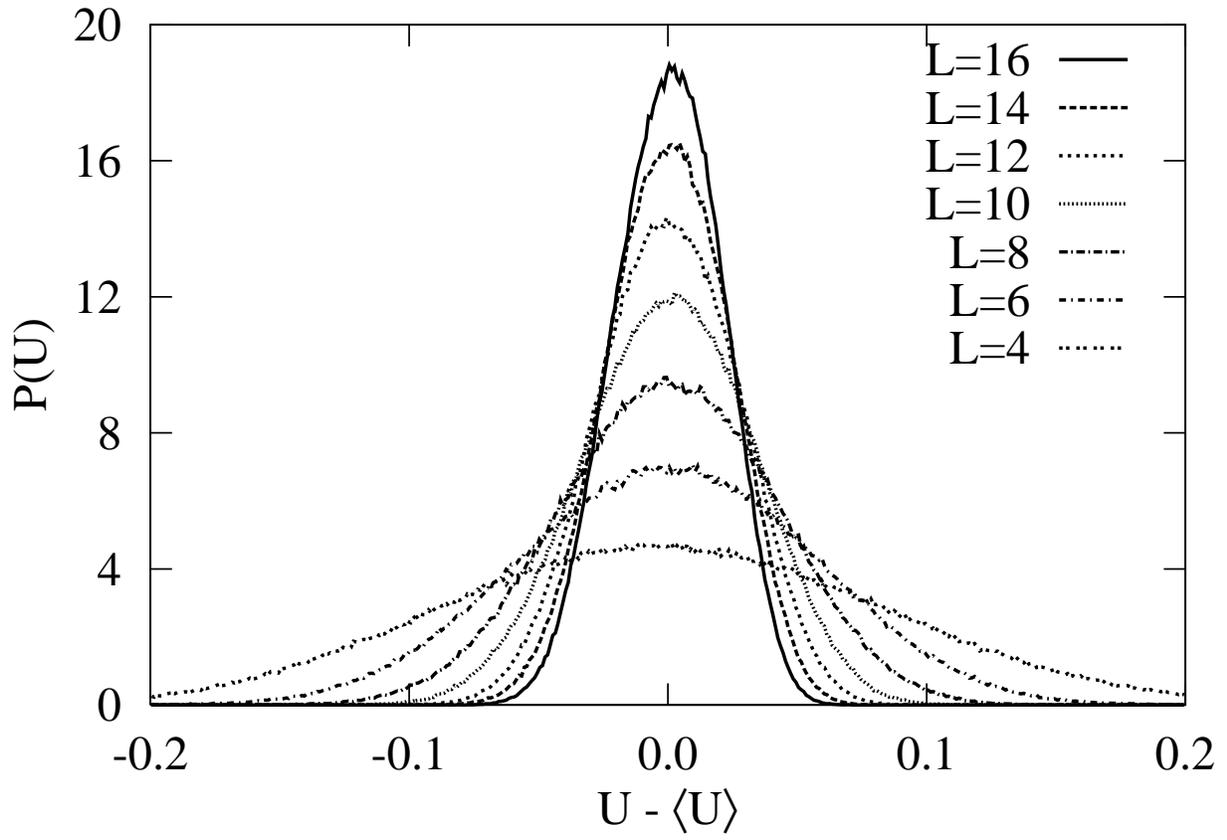}\\%
\end{center}
\caption{\label{fig:hist}MC probability distribution $P(U)$ for the
internal energy per site $U$ at the transition temperature
$T_{c}=0.681J/k_{\mathrm{B}}$ for several lattice sizes.}
\end{figure}

In Fig.\ \ref{fig:hist}, we show the probability distribution $P(U)$
of the internal energy per site $U$ for several lattice sizes at
$T_{c}$.   For each lattice size, we see only a single peak, which
sharpens with increasing lattice size.  This result is also in
contrast to that of Ref.\ \cite{hwang}, where the single peak splits
into two peaks as the lattice size increases.  This persistent
single-peak behavior suggests a continuous phase transition, in
contrast to the first-order phase transition found in Ref.\ \cite{hwang}.

\begin{figure}[h]
\begin{center}
\includegraphics[width=\textwidth]{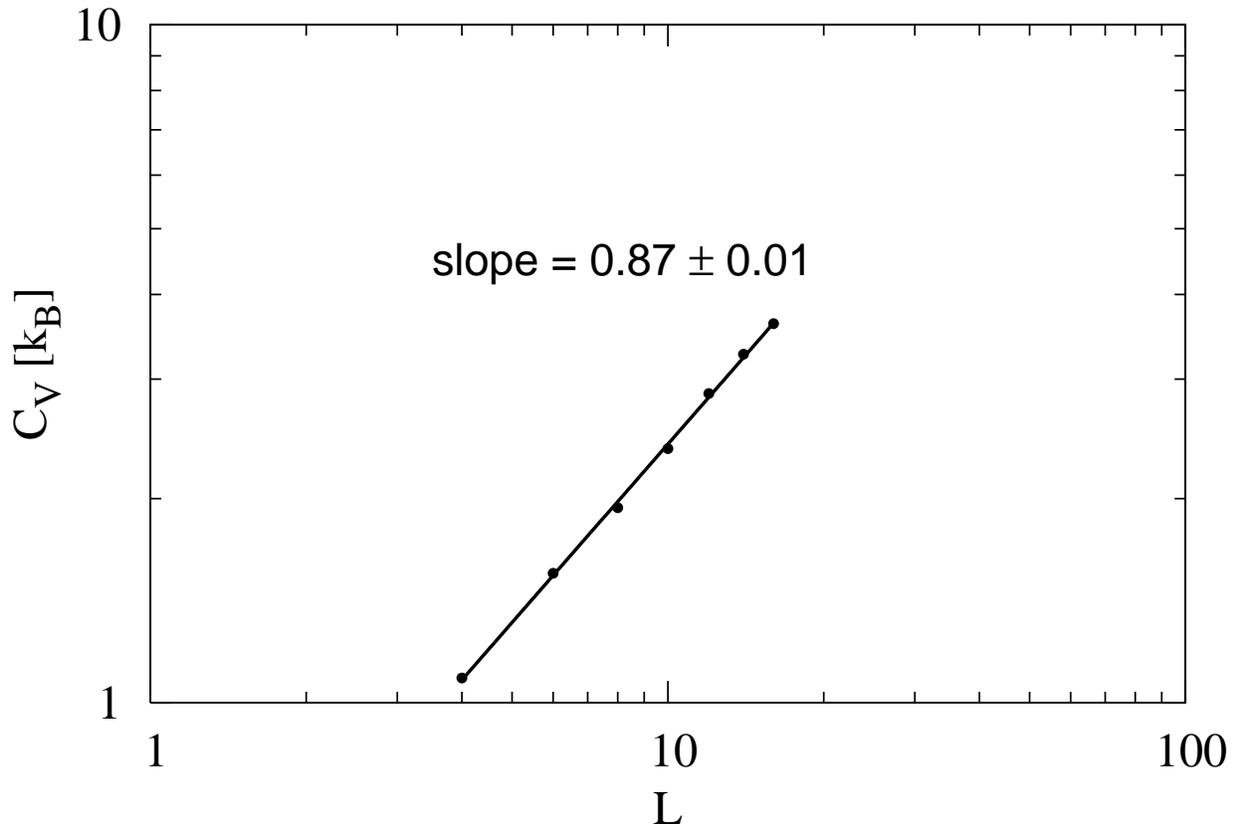}\\%
\end{center}
\caption{\label{fig:scaling_sh}Log-log plot of the maximum height
$C_V$ of the specific heat per site versus linear size $L$ of
the lattice.  The points are MC data; the full curve is the best-fit
line.   The slope of the fit line gives $\alpha/\nu=0.87\pm 0.01$.}
\end{figure}

The data shown in Fig.\ \ref{fig:sh} can be used in a standard way
to obtain an estimate of the critical exponent ratio $\alpha/\nu$.
From Eqs.\ (\ref{eq:xi}) and (\ref{eq:cv}), the maximum value of
$C_V$ for a lattice of edge $L$ is
\begin{equation}
C_{V}^{\max}\sim \xi^{\alpha/\nu} \sim L^{\alpha/\nu}.
\end{equation}
In Fig.\ \ref{fig:scaling_sh}, we plot $\log C_V^{\max}$ versus
$\log L$ at $T = T_c$.  The data is well described by a straight line.
The slope of this fitted straight line gives $\alpha/\nu=0.87\pm
0.01$. The present result is in contrast to the linear relation
between $C_{V}$ and $\log L$, corresponding to $\alpha = 0$
(logarithmic divergence) seen in the fully frustrated 2D XY model
as in Ref.\ \cite{teitel1}.

\begin{figure}[h]
\begin{center}
\includegraphics[width=\textwidth]{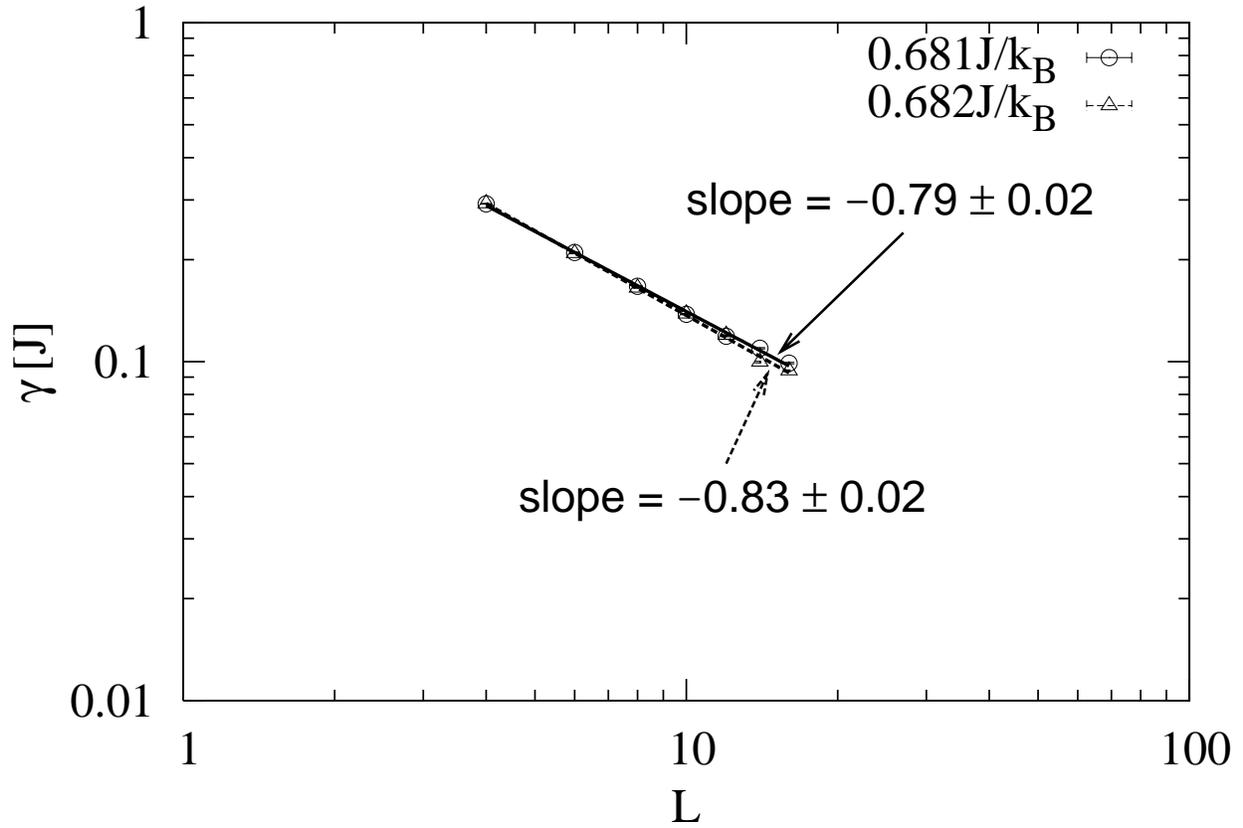}\\%
\end{center}
\caption{\label{fig:scaling_v_nu}Log-log plot of the helicity modulus
$\gamma$ versus linear size $L$ of the lattice.  The points are MC
data at two different temperatures $T=0.681J/k_{\mathrm{B}}$ and
$T=0.682J/k_{\mathrm{B}}$; the full curves are the best-fit lines.
The slopes of the fit lines give $v/\nu=0.79\pm 0.02$ at
$T=0.681J/k_{\mathrm{B}}$ and $v/\nu=0.83\pm 0.02$ at
$T=0.682J/k_{\mathrm{B}}$. The error bars from the jackknife
method are smaller than the symbol sizes.}
\end{figure}

\begin{figure}[h]
\begin{center}
\includegraphics[width=\textwidth]{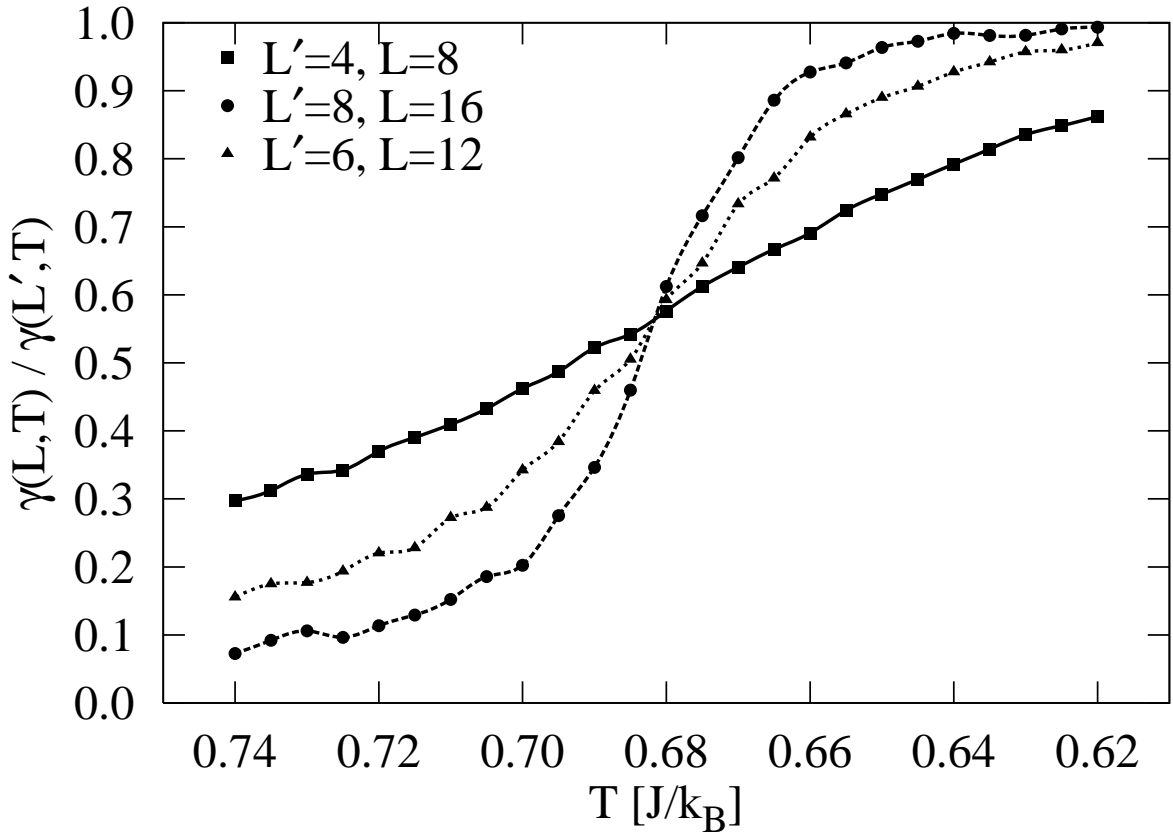}\\%
\end{center}
\caption{\label{fig:scaling_gamma}PRG method to extract the
transition temperature $T_{c}$ and the critical exponent $v/\nu$ of
the helicity modulus $\gamma$.   All three curves show $\gamma(L,
t)/\gamma(L^\prime, t)$ for the same ratio of $L/L^\prime=2$.  The
$x$ and $y$ coordinates of the intersection of the three curves
yield the values of $T_{c}$ and $(L/L^\prime)^{-v/\nu}$.}
\end{figure}

As we did for the specific heat $C_{V}$, we see from Eqs.\ (\ref{eq:xi})
and (\ref{eq:gamma}), or just from Eq.\ (\ref{eq:gammasf}), that the
helicity modulus $\gamma$ depends on the lattice size $L$ as
\begin{equation}
\gamma\sim \xi^{-v/\nu} \sim L^{-v/\nu}.
\end{equation}
In Fig.\ \ref{fig:scaling_v_nu}, we plot $\log \gamma$ versus $\log L$
at two different temperatures $T=0.681J/k_{\mathrm{B}}$ and
$T=0.682J/k_{\mathrm{B}}$. From the fits to the data, we get
$v/\nu=0.79\pm 0.02$ at $T=0.681J/k_{\mathrm{B}}$
and $v/\nu=0.83\pm 0.02$ at $T=0.682J/k_{\mathrm{B}}$. The error bars
from the jackknife method are also shown, but they are smaller than
the symbol sizes. The value of $v/\nu$ is very sensitive to small
changes in the assumed $T_c$. So we need a more reliable method to
calculate $v/\nu$. In Fig.\ \ref{fig:scaling_gamma}, we show the
results of a PRG study of the helicity modulus $\gamma(L, t)$.
As described near Eq.\ (\ref{eq:prg}), we plot $P_\gamma(L, L^\prime, t)$
for the fixed ratio $L/L^\prime = 2$, but for three different pairs
of sizes $L$ and $L^\prime$.  From the intersection of these three
$P_{\gamma}(L,L^\prime,t)$ curves as a function of $t$, we find
$T_{c}=0.681\pm 0.001J/k_{\mathrm{B}}$ and $v/\nu=0.82\pm 0.01$.

The previous two methods give only the ratios of two critical
exponents.  In order to find all three critical exponents, we need a
third method to extract $\nu$.   For this, we use another
renormalization group transformation \cite{newman5}.  We first
calculate $U(T)$  for a $16\times 16\times 16$ lattice.  Next, we
reduce each linear lattice dimension by a factor of $2$, by
combining a cube of eight adjacent sites into one; this is known as
a {\it blocking scheme} of the real-space renormalization group method.
This factor $2$ corresponds to the scaling factor $b$.
The phase of the merged eight sites is assigned by a simple additive
rule --- that is, it is taken as
\begin{equation}
\phi_{\mathrm{b}}=\arctan \left( \frac{\sum_{i}^{8}\sin \phi_{i}}
{\sum_{i}^{8}\cos \phi_{i}}\right),
\end{equation}
where  $-\pi \leq \phi_{\mathrm{b}}\leq\pi$. We add $2\pi$ when
$\phi_{\mathrm{b}} <0$ so that $\phi_{\mathrm{b}}$ satisfies
$0 \leq \phi_{\mathrm{b}}<2\pi$.
The correlation length $\xi^\prime$ of the blocked lattice is also
reduced by the scaling factor $b$, i.e.
\begin{equation}
\xi^\prime=\frac{\xi}{b} \label{eq:bxi}
\end{equation}
when $\xi^\prime$ is measured in terms of the lattice constant. But
since the correlation length should diverge at $T_c$, it follows that
\begin{equation}
\xi(T_c) =\xi^\prime (T_c).
\end{equation}

To use this scheme to obtain $\nu$, we also calculate the internal
energy per site $U^\prime$ of the rescaled $8\times 8\times 8$
lattice as a function of temperature $T$. This rescaled lattice
should behave very similarly to an $8\times 8\times 8$ lattice at a
different temperature $T^\prime$; so
\begin{equation}
U^\prime(T)=U(T^\prime).
\end{equation}
If we rewrite $T^\prime$ in terms of $T$, we obtain
\begin{equation}
T^\prime=U^{-1}[U^\prime(T)],
\end{equation}
where $U^{-1}$ is the functional inverse of the function $U$.
This relation between $T$ and $T^\prime$ is the desired
renormalization group transformation. As in Eq.\ (\ref{eq:xi}), the
rescaled correlation length $\xi^\prime$ can be expressed in terms
of the reduced temperature $t^\prime$ as
\begin{equation}
\xi^\prime\sim |t^\prime|^{-\nu}. \label{eq:bxiapp}
\end{equation}
From Eqs.\ (\ref{eq:xi}), (\ref{eq:bxi}), and (\ref{eq:bxiapp}),
we obtain
\begin{equation}
\left(\frac{t}{t^\prime}\right)^{-\nu}=b.
\label{eq:ratiot}
\end{equation}
Therefore, knowledge of the renormalization group transformation
taking the system from $T$ to $T^\prime$ gives the value of $\nu$.

Since Eqs.\ (\ref{eq:bxi}) and (\ref{eq:bxiapp}) are the asymptotic
forms near $T_{c}$, we can linearize the renormalization group
transformation using a Taylor series expansion near $T_{c}$ to
obtain
\begin{equation}
T^\prime-T_{c}=(T-T_{c})\left.\frac{\mathrm{d}T^\prime}{\mathrm{d}T}\right|_{T_{c}},
\end{equation}
or equivalently
\begin{equation}
\frac{t^\prime}{t}= \left.\frac{\mathrm{d}T^\prime}{\mathrm{d}T}\right|_{T_{c}}.
\label{eq:tratio}
\end{equation}
Inserting this result into Eq.\ (\ref{eq:ratiot}) yields
\begin{equation}
\nu=\frac{\log b}{\log \left.\frac{\mathrm{d}T^\prime}{\mathrm{d}T}\right|_{T_{c}}}.
\label{eq:nu}
\end{equation}

\begin{figure}[h]
\begin{center}
\includegraphics[width=\textwidth]{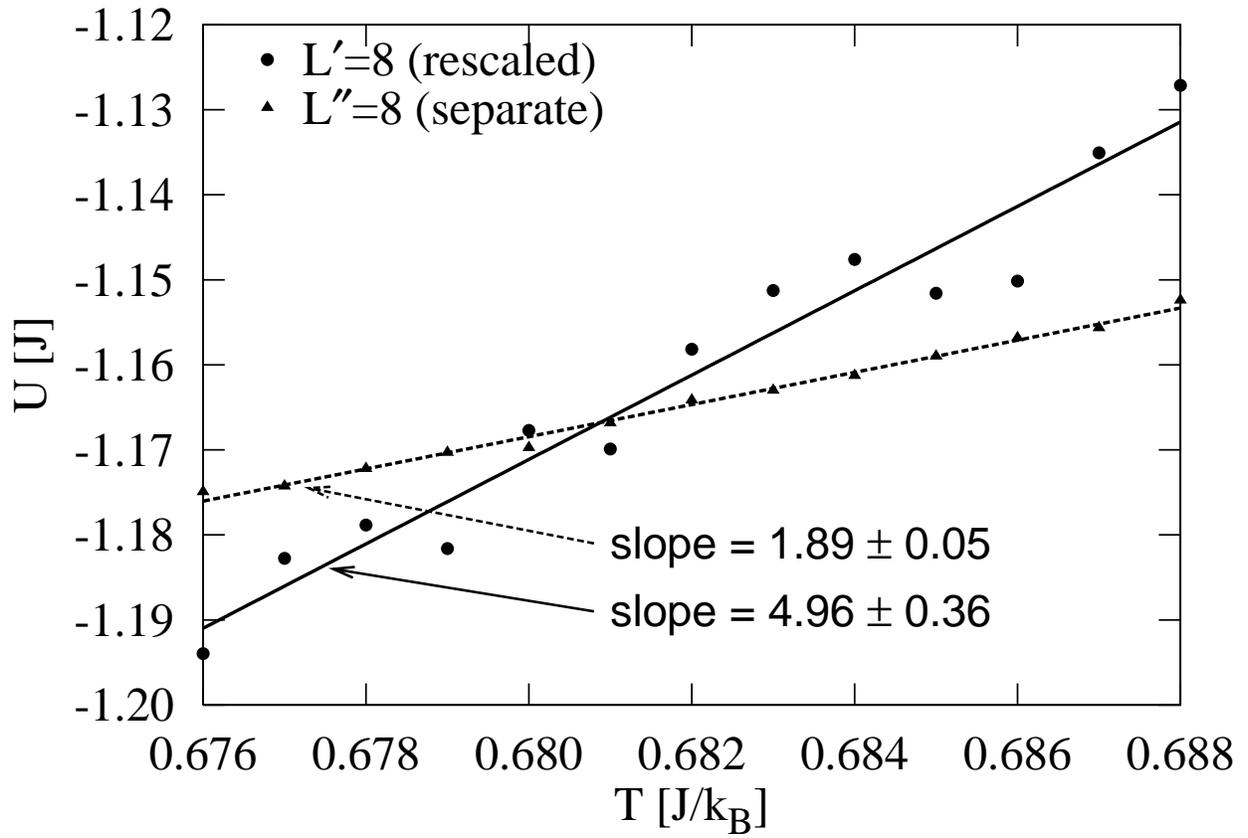}\\%
\end{center}
\caption{\label{fig:scaling_nu}RGT method to extract the transition
temperature $T_{c}$ and the critical exponent $\nu$ of the
correlation length.  The ratio of the slopes of the two fitting
lines at $T_{c}$ gives the value of $\nu$, according to Eq.\
(\ref{eq:nu}).}
\end{figure}

In Fig.\ \ref{fig:scaling_nu}, we show simulation results for the
rescaled $8\times 8\times 8$ lattice and for a separate unrescaled
lattice of the same size.   If we fit the two sets of points near
the critical temperature $T_{c}$ to two straight lines, the ratio of
the slopes yields the value of $(\mathrm{d}T^\prime/\mathrm{d}T)_{T_{c}}$,
according to Eq.\ (\ref{eq:tratio}).   We can then get $\nu$ from
Eq.\ (\ref{eq:nu}).  From the plot shown in Fig.\ \ref{fig:scaling_nu},
we obtain $\nu=0.72\pm 0.07$.
Since we already know $\alpha/\nu$ and $v/\nu$, we can use this
value of $\nu$ to obtain $\alpha=0.63\pm 0.07$ and $v=0.59\pm 0.07$.
In practice, the plot for the rescaled data has considerable uncertainty
in the slope; so $\nu$ cannot be determined with great accuracy using this
method, at least using simulations of the size we have carried out here.

All of our numerical data appears to be consistent with a single
phase transition at $T_{c}=0.681J/k_{\mathrm{B}}$.   By contrast,
there are two separate phase transitions in two-dimensional fully
frustrated XY model on a square or triangular lattice, as discussed
in Refs.\ \cite{olsson} and \cite{korshunov}. One of these is a KT
transition while the other is in the Ising universality class.
Although the two transition temperatures are close to each other,
the Ising transition temperature is slightly higher than that of the
KT transition.   Reference\ \cite{korshunov} explains the sequence of 
these phase transitions in terms of the loss of phase coupling 
across a domain wall.  Although our results are consistent with a 
single phase transition, we do not have sufficient data to rule out the
possibility that there could be two separate phase transitions in
our model, as further discussed below.

\begin{figure}[h]
\begin{center}
\includegraphics[width=\textwidth]{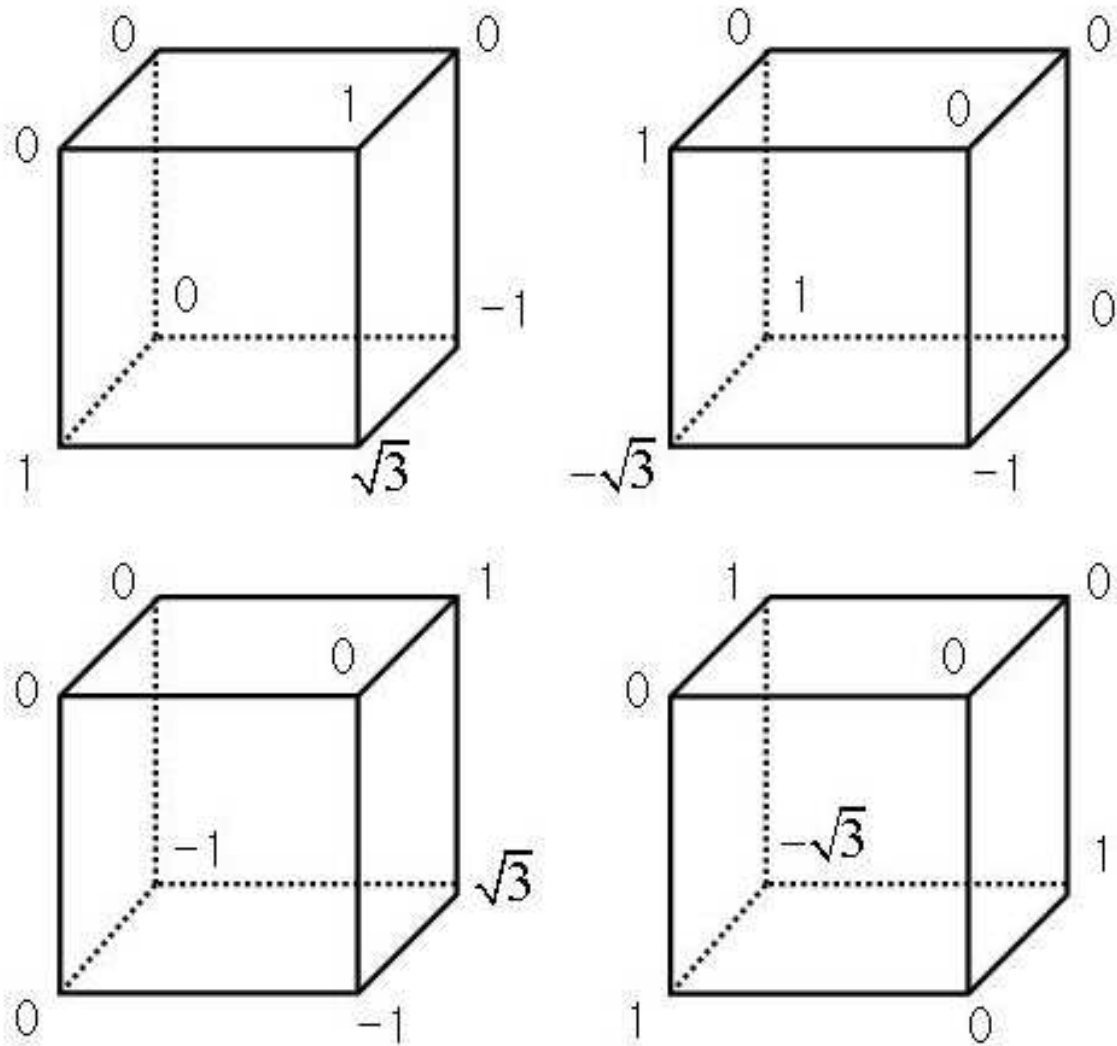}\\%
\end{center}
\caption{\label{fig:eigenmode}Four degenerate eigenmodes for the $2
\times 2 \times 2$ unit cell of the ordered state with ${\bf
f}=(1/2,1/2,1/2)$, corresponding to the mean-field transition
temperature $T_{c}^{\mathrm{MF}}=\sqrt{3}/2\ J/k_{\mathrm{B}}$.  The
numbers represent the order parameters $\eta_i$ of the mode, as
indicated in Eq.\ (\ref{eq:mfe}).  The phases reside on the nodes of the
lattice.}
\end{figure}

To gain further insight into our numerical results, we have also
used the mean-field approximation developed by Shih and Stroud
\cite{shih2} to find the relevant eigenmodes near the phase
transition at $T_{c}$. Near $T_{c}$, the linearized mean-field
equations are
\begin{equation}
\eta_{i}-\frac{\beta}{2}\sum_{j}J_{ij}e^{iA_{ij}}\eta_{j}=0,
\label{eq:mf}
\end{equation}
where $\eta_{i}\equiv \langle e^{i\phi_{i}}\rangle$ and
$\beta=1/k_{\mathrm{B}}T$. If we assume $J_{ij}=J$ and use
$T^\prime=k_{\mathrm{B}}T/J$, Eq.\ (\ref{eq:mf}) becomes
\begin{equation}
\eta_{i}-\frac{1}{2T^\prime}\sum_{j}e^{iA_{ij}}\eta_{j}=0.
\label{eq:mfe}
\end{equation}
In the mean-field approximation, $T_{c}$ is the highest value of $T$
such that Eq.\ (\ref{eq:mfe}) has a nontrivial solution. With our
frustration ${\bf f}=(1/2,1/2,1/2)$, the Hamiltonian is periodic
with a unit cell of $2 \times 2 \times 2$, which implies that Eq.\ 
(\ref{eq:mfe}) is a set of eight coupled homogeneous linear equations.
$T_c$ is given by the condition that the determinant of the matrix
of coefficients should vanish.  This requirement leads to two values
of $T^\prime=\pm \sqrt{3}/2 \sim 0.866$; each is fourfold
degenerate.   Thus, the transition temperature in the mean-field
approximation is $T_{c}^{\mathrm{MF}}=0.866J/k_{\mathrm{B}}$.  The
four degenerate eigenmodes corresponding to this eigenvalue are
$(0,-1,1,\sqrt{3},0,0,0,1)$, $(1,0,-\sqrt{3},-1,0,0,1,0)$,
$(-1,\sqrt{3},0,-1,0,1,0,0)$, and $(-\sqrt{3},1,1,0,1,0,0,0)$; these
modes are shown in Fig.\ \ref{fig:eigenmode}.  Of course, any linear
combination of these modes would also be an eigenmode.
Since the four eigenmodes are degenerate, they must be related by
some discrete symmetry operations of the lattice.  Thus, we might
expect that there are several degenerate ground states. Indeed, our
simulations at very low temperatures ($T = 0.01J/k_{\mathrm{B}}$)
do suggest that the system can readily fluctuate among several
different states at such temperatures.

\section{\label{sec:level4}Discussion}

We have investigated the phase transition in a fully frustrated 3D
XY model on a simple cubic lattice, corresponding to an applied
magnetic field ${\bf H}=(\Phi_{0}/a^2)(1/2,1/2,1/2)$.
In contrast to the case ${\bf f}=(1/3,1/3,1/3)$ as in Ref.\
\cite{hwang}, we see a {\em continuous} phase transition. We also
extract the critical exponent ratios $\alpha/\nu=0.87\pm 0.01$,
$v/\nu=0.82\pm 0.01$, and, with less accuracy, $\nu$ itself,
using a variety of numerical techniques.  We get $\alpha=0.63\pm 0.07$,
$v=0.59\pm 0.07$, and $\nu=0.72\pm 0.07$.

It is of some interest to compare these values with those of other
models.  In the isotropic (unfrustrated) 3D XY model, $\nu \sim
0.66-0.67$ \cite{li1,gottlob,janke,fernandez,schultka,nguyen1,
ryu,nho,krech,vestergren}, $\alpha \approx 0$ \cite{gottlob,janke,ryu}
($\alpha=-0.017$ in Ref.\ \cite{schultka}), and $v\simeq\nu$
\cite{li1,gottlob,ryu,nguyen1}. For the {\em anisotropic} but
unfrustrated XY model in $d = 3$, Ref.\ \cite{nguyen2} reported
that $\alpha \sim -0.007$. In the weakly frustrated XY model in
$d = 3$ (${\bf f} = (0, 0, f)$, with $f \leq 1/12$), it has been
reported that $\nu \approx 1.5$ \cite{nguyen2}. Reference\ \cite{kawamura1}
found that $\nu = 2.2 \pm 0.4$ in a random-coupling 3D XY model
with free boundary conditions and ${\bf f} = (0, 0, 1/2\pi)$, while
Ref.\ \cite{kawamura2} found that $\nu = 1.1 \pm 0.2$ in a
random-coupling 3D XY model with periodic boundary conditions
and ${\bf f} = (0, 0, 1/4)$. Our value of $\nu$ satisfies the
usual trend that $\nu$ becomes larger than $0.66-0.67$ when the
magnetic field is nonzero. However, our results appear not to
satisfy the two hyperscaling laws $\alpha = 2 - d\nu$ and
$v = (d - 2)\nu$. Possibly the reason is that at this phase
transition, two order parameters go to zero: the helicity modulus,
and a discrete, Ising-like order parameter related to the
amplitude of the structure factor at a characteristic wave vector.

The $T_{c}$ we obtain from the PRG method for $\gamma$
agrees very well with that found from the RGT method, and quite well
also with that estimated from $C_V(T)$.  It is, however, about $20\%$
lower than that obtained by the mean-field approximation.  This
deviation is not a surprise, because the Monte Carlo $T_c$ has been
shown to be lower than that found by mean-field theory in other
frustrated XY systems \cite{shih1}.

In order to avoid the problem of critical slowing down, instead of
just increasing the number of Monte Carlo steps, one might attempt
to use the cluster flipping algorithm, as first proposed by Swendsen
and Wang \cite{swendsen}, and later by Wolff \cite{wolff}, and many
others \cite{edwards,niedermayer,kandel,kawashima,cataudella2,machta}.
This algorithm is very successful in reducing critical slowing down
for unfrustrated models such as the $O(n)$ $\sigma$ model and the
Potts model.  But when the system is frustrated, especially fully
frustrated, a mere extension of the cluster algorithm does not
reduce the critical slowing down, and may even {\em increase} the
correlation time \cite{cataudella}. The reason is that the cluster
percolation temperature $T_{p}$ is much larger than $T_c$ in the
frustrated system, rendering the cluster flipping trivial at $T_{c}$
because the percolating cluster takes up almost the entire system.
Therefore, the cluster should be generated in such a way that
$T_{p}$ is closer to $T_{c}$ \cite{cataudella,cataudella2}. Critical
slowing down might also be reduced if the standard Metropolis
algorithm is combined with the cluster algorithm, resulting in a
hybrid algorithm \cite{nho,peczak,krech}.

The renormalization group transformation method
is based on certain assumptions which may lead to systematic
errors \cite{newman5}.   Specifically, it assumes that the blocked
system has a typical phase configuration of another 3D XY model on a
simple cubic lattice with $L^\prime=L/b$, i.e., that they appear
with the correct Boltzmann probabilities as the original states.
This assumption is not exactly correct, and contributes to some
systematic error, which cannot be easily estimated.  In the present
paper, we have tried to estimate these errors using the jackknife
method.   The renormalization group method is also affected by
finite-size effects,  because the original system has a different
size than does the rescaled system. To optimize the benefits of
this method, therefore, we should ideally run simulations on as
large a system as possible.  We should also do an extra simulation
on a system whose size is the same as that of the rescaled system,
and compare the two results.

We also comment briefly on the difference between our work and that
of Diep {\it et al.} \cite{diep}.  These authors did not compute the
helicity modulus, nor did they comment on the connection between the
model and a superconducting array in a magnetic field.  In addition,
because of the several techniques described above, we are able to
get more accurate values of the ratio $\alpha/\nu$, as well as
values for $v/\nu$ and of $\nu$ itself.

\begin{figure}[h]
\begin{center}
\includegraphics[width=\textwidth]{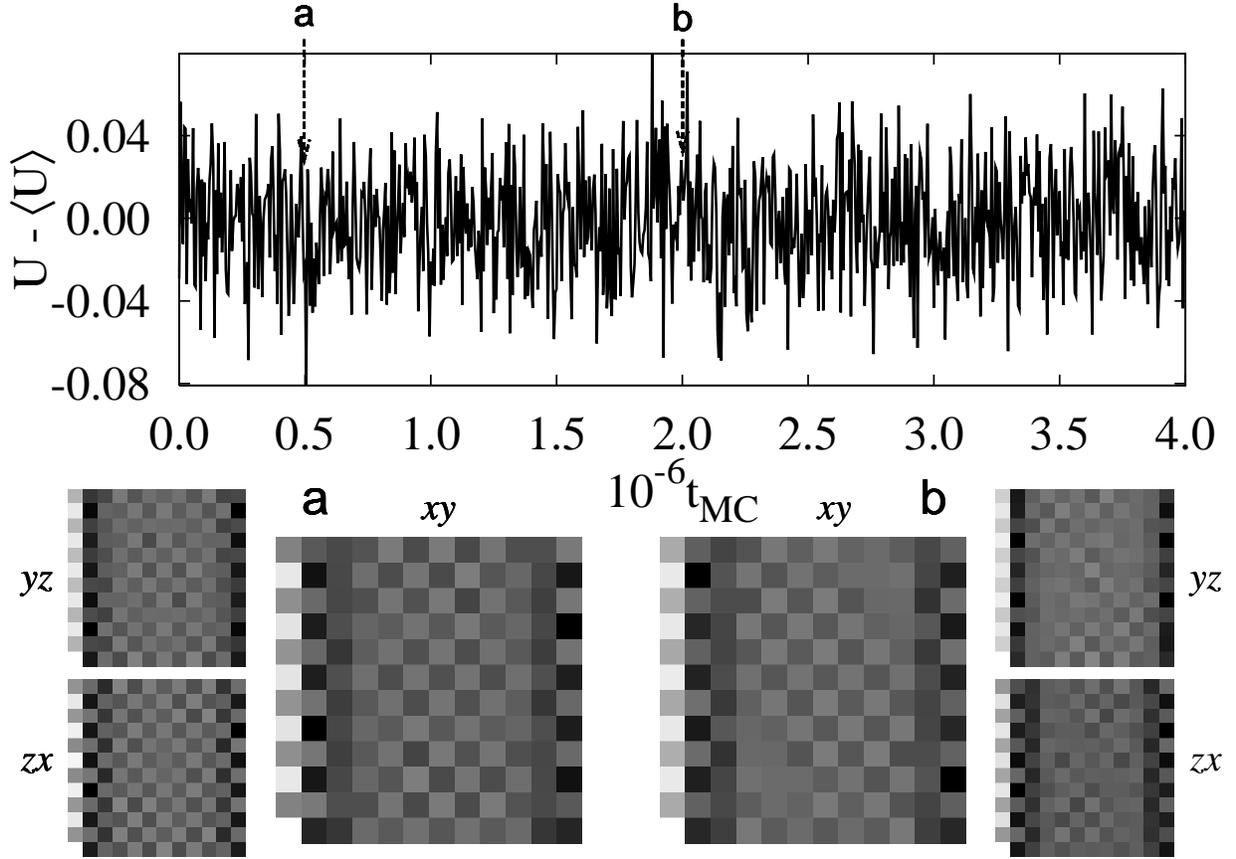}\\%
\end{center}
\caption{\label{fig:densitytc}Internal energy per site for a
lattice of size $12\times 12\times 12$ as a function of MC time, at
the transition temperature $T_{c}=0.681J/k_{\mathrm{B}}$. The
intensity plots $a$ and $b$ represent the vortex density-density
correlation functions $g_{zz}(x,y,L_{z}/2)$, $g_{xx}(L_{x}/2,y,z)$,
and $g_{yy}(x,L_{y}/2,z)$ denoted by $xy$, $yz$, and $zx$,
respectively, at the times $a$ and $b$ in the energy evolution
curve. A lighter color represents a larger value of the correlation
function.}
\end{figure}

\begin{figure}[h]
\begin{center}
\includegraphics[width=\textwidth]{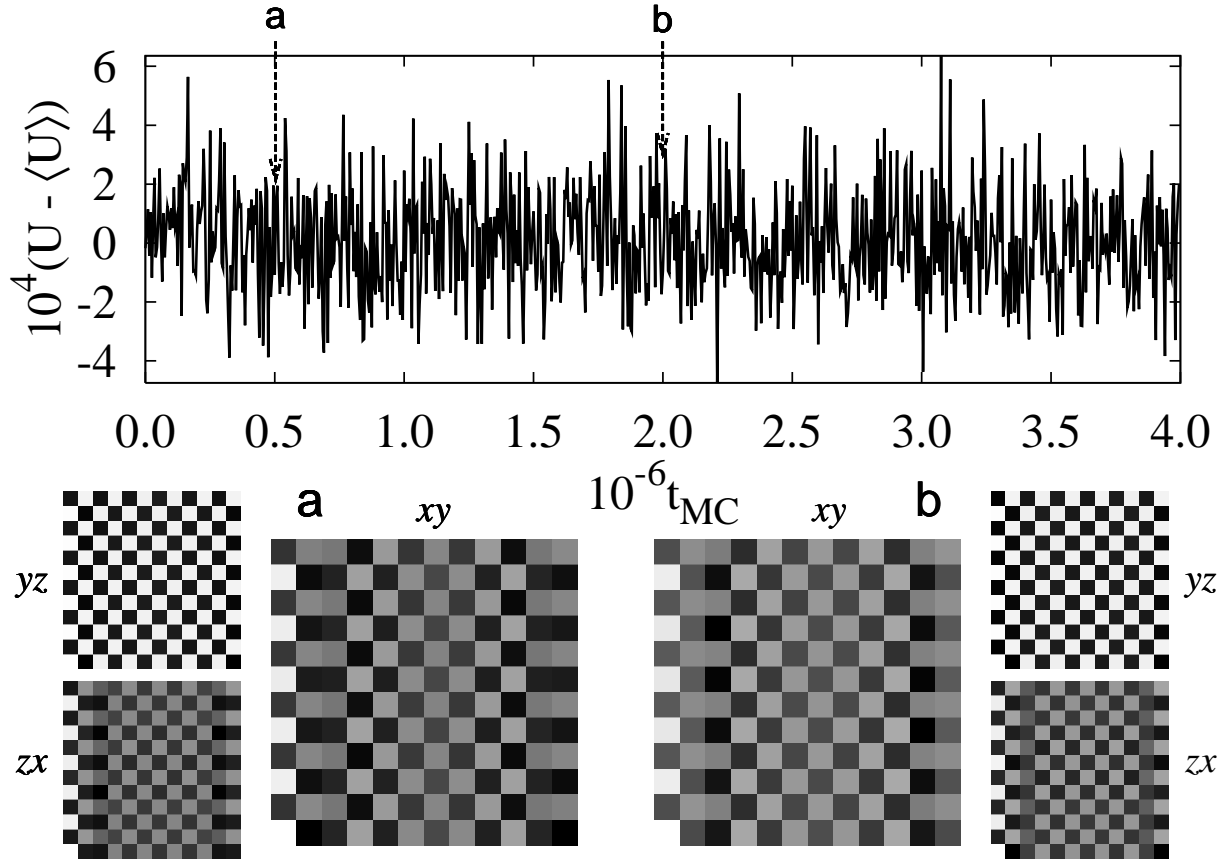}\\%
\end{center}
\caption{\label{fig:densitylowt}Same as Fig.\ \ref{fig:densitytc},
except that the lattice is at a very low temperature $T=0.01
J/k_{\mathrm{B}}$.}
\end{figure}

\begin{figure}[h]
\begin{center}
\includegraphics[width=\textwidth]{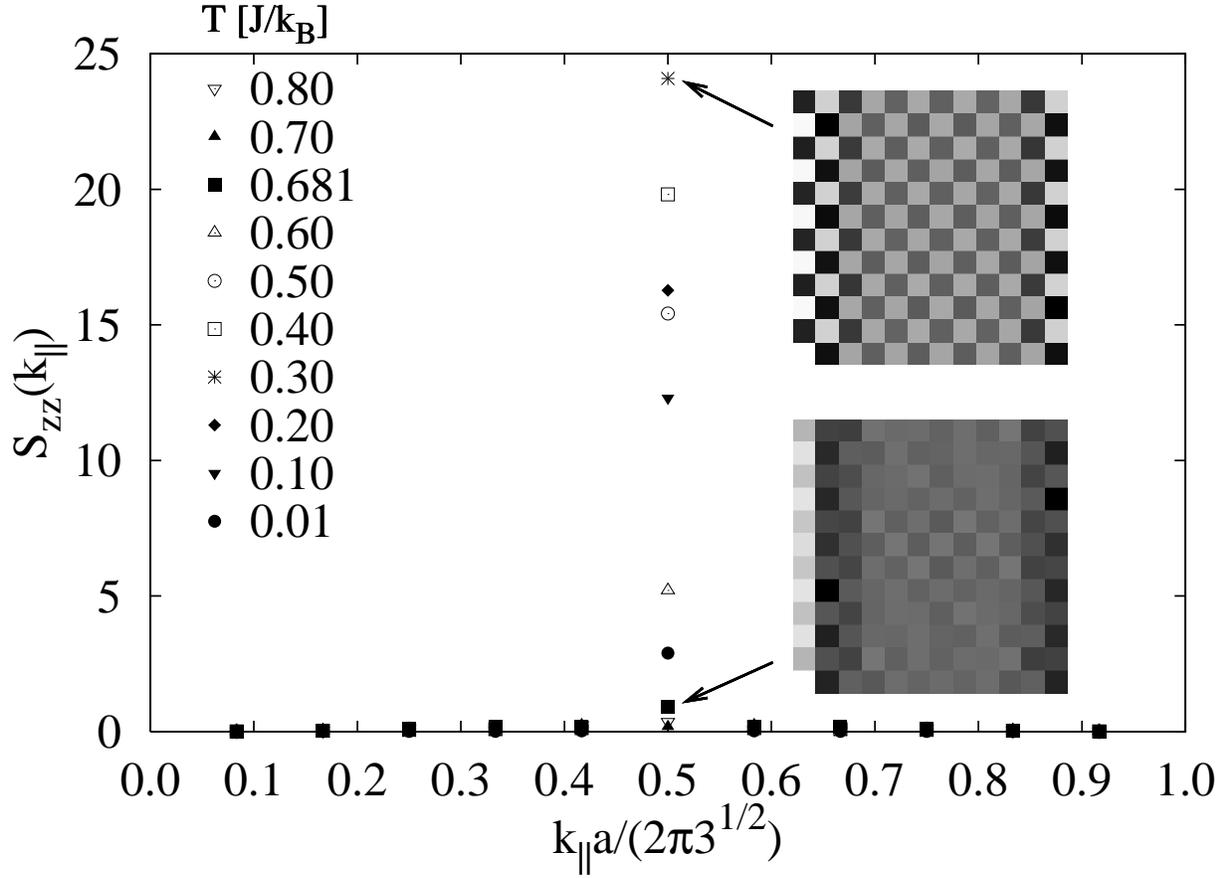}\\%
\end{center}
\caption{\label{fig:densitysf}Vortex structure factor
$S_{zz}(k_{\parallel})$ parallel to the applied magnetic field at
several temperatures for a lattice of size $12\times 12\times 12$
and periodic boundary conditions. The insets are the vortex
density-density correlation functions $g_{zz}(x,y,L_{z}/2)$ at two
different temperatures $T=0.30J/k_{\mathrm{B}}$ (top) and
$T=0.681J/k_{\mathrm{B}}$. A lighter color represents a larger value
of the correlation function.}
\end{figure}

Finally, we briefly discuss the possibility that the phase
transition at $T_c$ might actually be two separate phase
transitions.  In the 2D fully frustrated XY model, as already
mentioned, there are indeed two separate phase transitions: a KT
transition at a lower temperature, followed by an Ising-like
transition at a slightly higher temperature \cite{olsson,korshunov}
between a state in which the vortices are ordered in a checkerboard
pattern and a disordered vortex state. In the present case, the
transition at $T_c$ also involves the nearly simultaneous
disappearance of 3D XY order (signaled by the vanishing of the
helicity modulus) and a discrete order parameter (indicated by the
vanishing of the vortex structure factor).  Once again, in 3D, the
discrete order is characterized by a checkerboard configuration of
the vortex state below $T_c$ (see Figs.\
\ref{fig:densitytc}--\ref{fig:densitysf}), although in this case the
discrete order parameter is described by four degenerate modes, as
determined by the mean-field solution, rather than two as in the 2D
fully frustrated XY model \cite{choi2}.

Although the discrete and XY order appear to vanish at the same
temperature in 3D, and there is no evidence of two separate phase
transitions, we believe that our numerical results are not
sufficient to conclusively rule out two separate phase transitions.
In particular, we have not carried out careful numerical studies of
the critical behavior of the {\em discrete} order parameter. It
would be of interest to carry out further numerical studies,
especially of the discrete order parameter, to answer this question
definitively.

To summarize, we have studied the phase transition in the fully
frustrated XY model, using Monte Carlo simulations in conjunction
with two types of real-space renormalization group approaches. We
find, in agreement with previous work, that the phase transition is
continuous, and we obtain accurate values of the critical exponents
$\alpha/\nu$ and $v/\nu$, and a slightly less accurate value for
$\nu$ itself. The phase transition could, in principle, be probed
experimentally in a suitable three-dimensional lattice of coupled
superconducting grains, and possibly also in an assembly of cold
atoms in an optical lattice.

\begin{acknowledgments}
This work was supported by NSF Grant No. DMR04-13395. All of the calculations
were carried out on the P4 Cluster at the Ohio Supercomputer Center,
with the help of a grant of time.
\end{acknowledgments}

\newpage

\begin{center}
TABLE
\end{center}

\vspace{0.1in}
\begin{tabular}{c|c}
\hline \hline \multicolumn{1}{c|}{$L$} &
\multicolumn{1}{c}{$\tau(t_{\mathrm{MC}})$} \\
\hline
4 & 14  \\
6 &  29  \\
8 &  53  \\
10 &  94  \\
12 &  148  \\
14 &  164  \\
16 &  352  \\
\hline \hline
\end{tabular}

\vspace{0.2in}

\noindent TABLE I: Calculated correlation time
$\tau$ for several lattice sizes $L$, evaluated at $T_c(L)$.
$\tau$ is measured in the units of MC steps per site.

\newpage

\begin{center}
FIGURES
\end{center}

\vspace{0.1in}

\newpage

\newpage

\newpage

\newpage

\newpage

\newpage

\newpage

\newpage

\newpage

\newpage

\newpage

\newpage

\end{document}